# Anomalous behavior of *q*-averages in nonextensive statistical mechanics

Sumiyoshi Abe[1,2,3]

[1] Department of Physical Engineering, Mie University, Mie 514-8507, Japan[4]

[2] Institut Supérieur des Matériaux et Mécaniques Avancés, 44 F. A. Bartholdi, 72000 Le Mans, France

[3] Inspire Institute Inc., McLean, Virginia 22101, USA

E-mail: suabe@sf6.so-net.ne.jp

**Abstract.** A generalized definition of average, termed the *q*-average, is widely employed in the field of nonextensive statistical mechanics. Recently, it has however been pointed out that such an average value may behave unphysical under specific deformations of probability distributions. Here, the following three issues are discussed and clarified. Firstly, the deformations considered are physical and may experimentally be realized. Secondly, in view of thermostatistics, the *q*-average is unstable in both finite and infinite discrete systems. Thirdly, a naive generalization of the discussion to continuous systems misses a point, and a norm better than the $L^1$-norm should be employed for measuring the distance between two probability distributions. Consequently, stability of the *q*-average is shown not to be established in all the cases.

___________________________________

[4] Permanent address.

## 1. Introduction

There exist a number of thermostatistical systems in nature, which are exotic from the traditional viewpoint of Boltzmann-Gibbs statistical mechanics. They often possess/exhibit broken ergodicity, strong correlations between elements leading to inseparability, nontrivial portraits of phase spaces or configuration spaces and long-range interactions, for example. In the past decade, nonextensive statistical mechanics [1, 2], which is a generalization of the Boltzmann-Gibbs theory, has been expected to offer a framework for describing the properties of such systems. The prevailing formulation of nonextensive statistical mechanics employs a generalized definition of average termed the $q$-average [see equation (5) below]. In this paper, we rigorously examine if such a generalization is possible.

Consider measurement of a physical quantity, $Q = \{Q_i\}_{i=1, 2, ..., W}$, of a thermostatistical system. To obtain reliable information on the probability distribution, the measurement has to be repeated. In reality, two probability distributions, $\{p_i\}_{i=1, 2, ..., W}$ and $\{p'_i\}_{i=1, 2, ..., W}$, thus obtained may slightly be different from each other. Here, $W$ stands for the number of accessible microstates and is a very large number, typically being $2^{10^{23}}$. Such a difference can be quantified, for example, by comparing the average values of $Q$ with respect to those probability distributions. Since $Q$ is supposed to behave well, the calculated average values are expected to be close to each other. This natural requirement is mathematically expressed by the following formal

predicate [3]:

$$(\forall \varepsilon > 0) \ (\exists \delta > 0) \ (\forall W) \ \left( \|p - p'\|_1 < \delta \ \Rightarrow \ |\langle Q \rangle[p] - \langle Q \rangle[p']| < \varepsilon \right) \quad (1)$$

for *any* pair of probability distributions, $\{p_i\}_{i=1, 2, ..., W}$ and $\{p'_i\}_{i=1, 2, ..., W}$, where

$$\|p - p'\|_1 = \sum_{i=1}^{W} |p_i - p'_i| \quad (2)$$

is the distance between these two probability distributions defined in terms of the $l^1$-norm, and $\langle Q \rangle[p]$ ($\langle Q \rangle[p']$) stands for the "average" of $Q$ with respect to $\{p_i\}_{i=1, 2, ..., W}$ ($\{p'_i\}_{i=1, 2, ..., W}$). Other kinds of norms could also be considered, but what is relevant to discrete systems is the present $l^1$-norm, which is able to make $\|p - p'\|_1$ independent of $W$ (see the later discussion). The average, $\langle Q \rangle$, is said to be *stable* or *robust*, if the condition in equation (1) is satisfied. This is somewhat analogous to Lesche's stability condition on entropic functionals [4]-[10] (see also [11] for a comment on [8]). Mathematically, it is concerned with uniform continuity of a functional.

In the scheme in equation (1), it is important to note that the condition on $W$ comes after those on $\varepsilon$ and $\delta$. This implies that the large-$W$ limit, for example, has to be taken at the end of calculation.

The standard definition of average, referred to as the normal average,

$$\langle Q \rangle_1[p] = \sum_{i=1}^{W} Q_i \, p_i \tag{3}$$

is clearly stable. This is simply seen as follows:

$$\left| \langle Q \rangle_1[p] - \langle Q \rangle_1[p'] \right| = \left| \sum_{i=1}^{W} Q_i \, (p_i - p'_i) \right|$$

$$\leq \sum_{i=1}^{W} |Q_i| |p_i - p'_i|$$

$$\leq |Q|_{\max} \cdot \| p - p' \|_1, \tag{4}$$

where $|Q|_{\max} \equiv \max\{|Q_i|\}_{i=1, 2, \ldots, W}$. Thus, there, in fact, exists $\delta$ such that $\delta = \varepsilon / |Q|_{\max}$. This is actually an obvious result since $\langle Q \rangle_1[p]$ is a bounded linear functional of $\{p_i\}_{i=1, 2, \ldots, W}$.

Now, in the field of nonextensive statistical mechanics [1, 2], a possibility is open for generalizing the definition of average. There, the so-called $q$-average

$$\langle Q \rangle_q[p] = \frac{\sum_{i=1}^{W} Q_i \, (p_i)^q}{\sum_{j=1}^{W} (p_j)^q} \qquad (q > 0), \tag{5}$$

is prevailing. This quantity is reduced to the normal average in equation (3) in the limit $q \to 1$. In general, it is a nonlinear functional of $\{p_i\}_{i=1, 2, \ldots, W}$, and this fact makes the problem of its stability nontrivial.

The purpose of this paper is to discuss and clarify that the $q$-average is unstable unless $q \to 1$. To explicitly do so, we consider specific deformations of probability distributions. Then, we present the following three results. Firstly, the deformations considered here are physical and may be realized in a laboratory. Secondly, in view of thermostatistics, stability of the $q$-average cannot be verified in both finite and infinite discrete systems. Then, thirdly, a naive generalization of the $l^1$-norm in equation (2) to the-$L^1$ norm for continuous systems misses an important physical point. Thus, we conclude that stability of the $q$-average cannot be established in all the cases, and accordingly the formulation of nonextensive statistical mechanics has to be amended.

## 2. Specific deformations of probability distributions and their physicality

To evaluate the quantity $\left| \langle Q \rangle_q [p] - \langle Q \rangle_q [p'] \right|$, let us examine the following specific deformations of the probability distributions:

i) for $0 < q < 1$;

$$p_i = \delta_{i1}, \quad p'_i = \left(1 - \frac{\delta}{2} \frac{W}{W-1}\right) p_i + \frac{\delta}{2} \frac{1}{W-1}, \tag{6}$$

ii) for $q > 1$;

$$p_i = \frac{1}{W-1}(1-\delta_{i1}), \quad p'_i = \left(1-\frac{\delta}{2}\right)p_i + \frac{\delta}{2}\delta_{i1}. \tag{7}$$

These have been considered in the context of stabilities of generalized entropies in [4, 5]. In both i) and ii), holds $\|p - p'\|_1 = \delta$, which is in fact independent of $W$, making stability analysis possible freely from the system size.

Here, we wish to mention that the deformations from $p_i$ to $p'_i$ in equations (6) and (7) are indeed physical and may even be realized in quantum-mechanical experiments. This is because as follows. Recall that the quantum $q$-average reads $\langle \hat{Q} \rangle_q [\hat{\rho}] = \mathrm{Tr}(\hat{Q}\hat{\rho}^q) / \mathrm{Tr}(\hat{\rho}^q)$, where $\hat{\rho}$ and $\hat{Q}$ are a density matrix and an observable, respectively. The deformed density matrix, $\hat{\rho}'$, corresponding to $p'_i$'s in equations (6) and (7) is given by the convex combination of the completely random state, $\hat{I}/W$ (with $\hat{I}$ being the identity matrix), and the normalized pure eigenstate, $|u_1\rangle\langle u_1|$ (corresponding to the first eigenvalue $Q_1$ of $\hat{Q}$) in $W$ dimensions:

$$\hat{\rho}' = \lambda \frac{\hat{I}}{W} + (1-\lambda)|u_1\rangle\langle u_1|, \tag{8}$$

where $0 \leq \lambda \leq 1$. $\lambda$ is given in terms of the fidelity, $F$, with respect to the reference state, $|u_1\rangle\langle u_1|$, as $\lambda = (1-F)W/(W-1)$, provided that $1/W \leq F \leq 1$. In case i), $\lambda = (\delta/2)W/(W-1)$, whereas $\lambda = (1-\delta/2)W/(W-1)$ in case ii). $\hat{\rho}'$ in equation

(8) is referred to as the Werner state [12] when $|u_1\rangle$ is maximally entangled. It is known [13, 14] that such a state can be generated for a bipartite system using spontaneous parametric down-conversion. For a large number of quanta, there is a technical difficulty in realizing a maximally entangled state. However, $|u_1\rangle$ does not have to be maximally entangled: any pure eigenstate is sufficient in our discussion, avoiding the main technical difficulty. In addition, we point out the fact [15] that the state in equation (8) is actually the thermal state

$$\hat{\rho}' = \frac{1}{Z(\beta)} e^{-\beta \hat{H}} \qquad (Z(\beta) = \mathrm{Tr}\, e^{-\beta \hat{H}}) \qquad (9)$$

with the projector Hamiltonian, $\hat{H} = -g |u_1\rangle\langle u_1|$, where $g$ is a positive constant having the dimension of temperature, $\beta^{-1}$. $\lambda$ is related to the inverse temperature $\beta$ and $g$ as $\lambda = W / (W + e^{\beta g} - 1)$.

3. **Discrete systems**

First, let us evaluate $\left| \langle Q \rangle_q [p] - \langle Q \rangle_q [p'] \right|$ for the deformations in equations (6) and (7). A straightforward calculation shows that, in case i),

$$\left| \langle Q \rangle_q [p] - \langle Q \rangle_q [p'] \right|$$

$$= \left| Q_1 - \frac{(1-\delta/2)^q Q_1 + (\delta/[2(W-1)])^q (W\overline{Q} - Q_1)}{(1-\delta/2)^q + (\delta/2)^q (W-1)^{1-q}} \right|, \qquad (10)$$

and, in case ii),

$$\left| \langle Q \rangle_q [p] - \langle Q \rangle_q [p'] \right|$$

$$= \left| \frac{W\overline{Q} - Q_1}{W-1} - \frac{(\delta/2)^q Q_1 + [(1-\delta/2)/(W-1)]^q (W\overline{Q} - Q_1)}{(\delta/2)^q + (1-\delta/2)^q (W-1)^{1-q}} \right|, \qquad (11)$$

where $\overline{Q} = (1/W) \sum_{i=1}^{W} Q_i$ is the arithmetic mean of $Q$.

In the limit $W \to \infty$, $\left| \langle Q \rangle_q [p] - \langle Q \rangle_q [p'] \right|$ converges to $\left| \overline{Q} - Q_1 \right|$ in both cases i) and ii). Therefore, the condition in equation (1) is violated, and the $q$-average is unstable in such a limit.

It is worth mentioning that the limits $q \to 1$ and $W \to \infty$ do not commute, since the normal average is stable as shown in equation (4).

Next, let us evaluate equations (10) and (11) with $W$ that is finite but very large, typically being $W \sim 2^{10^{23}}$ in thermostatistical systems. In case i), in order for $\left| \langle Q \rangle_q [p] - \langle Q \rangle_q [p'] \right|$ to be small, it is necessary that $(\delta/2)^q W^{1-q} \ll (1-\delta/2)^q$ holds. This condition leads to

$$\delta << 2^{-10^{23}} \qquad (12)$$

for $q = 1/2$. Similarly, in case ii), should hold $(\delta/2)^q << (1-\delta/2)^q W^{1-q}$, which leads to

$$\delta << 2^{-10^{23}/3} \qquad (13)$$

for $q = 3/2$.

These values of $\delta$ are extremely small, and it is unlikely that realization of such overwhelmingly high precision is physically possible in measurements of probability distributions. Basically, it is reasonable to believe that if the result is different between $W \sim 2^{10^{23}}$ and $W \to \infty$, then the model itself must be pathological.

Consequently, in view of thermostatistics, the $q$-averages are unstable in both finite and infinite discrete systems.

## 4. Continuous systems

In a recent paper [16], it has been claimed that the $q$-averages are stable in continuous systems. The discussion given there is based on a naive generalization of the $l^1$-norm in discrete systems to the $L^1$-norm, with which the distance between two normalized probability densities, $f(x)$ and $f'(x)$, defined in the range $0 \leq x \leq 1$ is given by

$$\| f - f' \|_1 = \int_0^1 dx \, | f(x) - f'(x) |. \tag{14}$$

It is shown [16] that the $q$-average, $\langle Q \rangle_q [f] = \int_0^1 dx \, Q(x) [f(x)]^q / \int_0^1 dx' \, [f(x')]^q$, satisfies $\left| \langle Q \rangle_q [f] - \langle Q \rangle_q [f'] \right| < c \delta^\alpha$, where $\delta = \| f - f' \|_1$, and $c$ and $\alpha$ are positive constants.

However, the above discussion misses a point. The distance in equation (14) poorly describes "closeness" between $f(x)$ and $f'(x)$. To see it, let us look at the following simple deformation:

$$f(x) = 1, \quad f'(x) = \begin{cases} \dfrac{2}{1 - \sqrt{1 - 2\delta}} & \left( 0 \leq x \leq \left( \dfrac{1 - \sqrt{1 - 2\delta}}{2} \right)^2 \right) \\ \dfrac{2}{3 - \sqrt{1 - 2\delta}} & \left( \left( \dfrac{1 - \sqrt{1 - 2\delta}}{2} \right)^2 < x \leq 1 \right) \end{cases}. \tag{15}$$

The distance between them defined by the $L^1$-norm is calculated to be $\| f - f' \|_1 = \delta$, although they are actually quite distinct from each other.

In mathematics, better norms are known. One such example is [17]

$$\| f - f' \| = \sup_{0 \leq x \leq 1} | f(x) - f'(x) |. \tag{16}$$

Clearly, the following inequality holds:

$$\|f - f'\|_1 \leq \|f - f'\|, \qquad (17)$$

implying that the distance in equation (16) introduces, in the space of probability distributions on the unit interval, metric topology finer than that defined by the $L^1$-norm. In fact, equation (16) for $f(x)$ and $f'(x)$ in equation (15) yields

$$\|f - f'\| = \frac{1 - \delta + \sqrt{1 - 2\delta}}{\delta}, \qquad (18)$$

which is larger than $1/\delta$ if $\delta < \sqrt{2} - 1$. Thus, we see that stability of the $q$-averages cannot be established also in continuous systems.

## 5. Concluding remarks

We have discussed and clarified that stability of the $q$-averages cannot be established in both discrete and continuous systems. In addition, we have also pointed out that the specific deformations of the probability distributions considered in the discussion are quite physical and may be realized in an optical-physics laboratory.

In a recent work [18], it has been shown that the normal-average formalism of nonextensive statistical mechanics (and the references quoted therein) is consistent with the generalized *H*-theorem, whereas the *q*-average formalism is not. There are also discussions about a number of conceptual difficulties with the *q*-averages [19]. In addition, it is suggested in [20] that based on these observations the homogeneous entropy, instead of Tsallis' entropy, should be used for describing asymptotically power-law distributions. The problem of the definition of averages is one of the cruxes in nonextensive statistical mechanics. The present results combined with those presented in [3, 18] show that what to be employed in nonextensive statistical mechanics is the normal averages, and not the *q*-averages.

**Acknowledgment**

This work was supported in part by a Grant-in-Aid for Scientific Research from the Japan Society for the Promotion of Science.

**References**

[1]   Abe S and Okamoto Y (eds.), 2001 *Nonextensive Statistical Mechanics and Its*

*Applications* (Heidelberg: Springer-Verlag,)

[2] Tsallis C, 2009 *Introduction to Nonextensive Statistical Mechanics: Approaching a Complex World* (New York: Springer Science+Business Media)

[3] Abe S, 2008 *Europhys. Lett.* **84** 60006

[4] Lesche B, 1982 *J. Stat. Phys.* **27** 419

Lesche B, 2004 *Phys. Rev. E* **70** 017102

[5] Abe S, 2002 *Phys. Rev. E* **66** 046134

Abe S, 2004 *Physica D* **193** 84

Abe S, 2004 *Contin. Mech. Thermodyn.* **16** 237

[6] Naudts J, 2004 *Rev. Math. Phys.* **16** 809

[7] Abe S, Kaniadakis G and Scarfone A M, 2004 *J. Phys. A* **37** 10513

[8] Curado E M F and Nobre F D, 2004 *Physica A* **335** 94

[9] Abe S, Lesche B and Mund J, 2007 *J. Stat. Phys.* **128** 1189

[10] Ubriaco M R, 2009 *Preprint* 0902.2726

[11] El Kaabouchi A, Wang Q A, Ou C J, Chen J C, Su G Z and Le Méhauté A, 2009 *Preprint* 0903.4169

[12] Werner R F, 1989 *Phys. Rev. A* **40** 4277

[13] Zhang Y-S, Huang Y-F, Li C-F and Guo G-C, 2002 *Phys. Rev. A* **66** 062315

[14] Barbieri M, De Martini F, Di Nepi G, Mataloni P, D'Ariano G M and Macchiavello C, 2003 *Phys. Rev. Lett.* **91** 227901

[15] Abe S, Usha Devi A R and Rajagopal A K, 2008 *Preprint* 0803.1430

[16] Hanel R, Thurner S and Tsallis C, 2009 *Europhys. Lett.* **85** 20005

[17] Yosida K, 1971 *Functional Analysis* 3rd edition (Berlin: Springer-Verlag)

[18] Abe S, 2009 *Phys. Rev. E* **79** 041116

[19] Abe S, "Conceptual difficulties with the *q*-averages in nonextensive statistical mechanics", to appear in *Proceedings of Statistical Mechanics and Mathematics of Complex Systems*.

[20] Lutsko J F, Boon J P and Grosfils P, 2009 *Preprint* 0902.4579, to appear in *Europhys. Lett.*